\documentclass[aps,prl,twocolumn,superscriptaddress]{revtex4-1}
\pdfoutput=1
\usepackage{graphicx}
\usepackage{amsmath,amsfonts,amssymb}
\usepackage[colorlinks=true,linkcolor=blue,citecolor=blue]{hyperref}
\usepackage{siunitx}

\begin{document}
\title{Giant valley-isospin conductance oscillations in ballistic graphene}

\author{Clevin Handschin}
\thanks{These authors contributed equally}
\affiliation{Department of Physics, University of Basel, Klingelbergstrasse 82, CH-4056 Basel, Switzerland}

\author{P\'eter Makk}
\thanks{These authors contributed equally}
\email{Peter.makk@unibas.ch}
\affiliation{Department of Physics, University of Basel, Klingelbergstrasse 82, CH-4056 Basel, Switzerland}

\author{Peter Rickhaus}
\thanks{These authors contributed equally}
\affiliation{Department of Physics, University of Basel, Klingelbergstrasse 82, CH-4056 Basel, Switzerland}

\author{Romain Maurand}
\affiliation{CEA, INAC-PHELIQS, University Grenoble Alpes, F-3800 Grenoble, France}

\author{Kenji Watanabe}
\affiliation{National Institute for Material Science, 1-1 Namiki, Tsukuba, 305-0044, Japan\\}

\author{Takashi Taniguchi}
\affiliation{National Institute for Material Science, 1-1 Namiki, Tsukuba, 305-0044, Japan\\}

\author{Klaus Richter}
\affiliation{Institut für Theoretische Physik, Universitat Regensburg, D-93040 Regensburg, Germany}

\author{Ming-Hao Liu}
\affiliation{Institut für Theoretische Physik, Universitat Regensburg, D-93040 Regensburg, Germany}
\affiliation{Department of Physics, National Cheng Kung University, Tainan 70101, Taiwan}

\author{Christian Sch\"onenberger}
\email{Christian.Schoenenberger@unibas.ch}
\affiliation{Department of Physics, University of Basel, Klingelbergstrasse 82, CH-4056 Basel, Switzerland}

\begin{abstract}
At high magnetic fields the conductance of graphene is governed by the half-integer quantum Hall effect. By local electrostatic gating a \textit{p-n} junction perpendicular to the graphene edges can be formed, along which quantum Hall channels co-propagate. It has been predicted by Tworzid\l{}o and co-workers that if only the lowest Landau level is filled on both sides of the junction, the conductance is  determined by the valley (isospin) polarization at the edges and by the width of the flake. This effect remained hidden so far due to scattering between the channels co-propagating along the \textit{p-n} interface (equilibration). Here we investigate \textit{p-n} junctions in encapsulated graphene with a movable \textit{p-n} interface with which we are able to probe the edge-configuration of graphene flakes. We observe large quantum conductance oscillations on the order of \si{e^2/h} which solely depend on the \textit{p-n} junction position providing the first signature of isospin-defined conductance. Our experiments are underlined by quantum transport calculations.
\end{abstract}

\maketitle

\begin{figure}[tb]
    \centering
      \includegraphics[width=1\columnwidth]{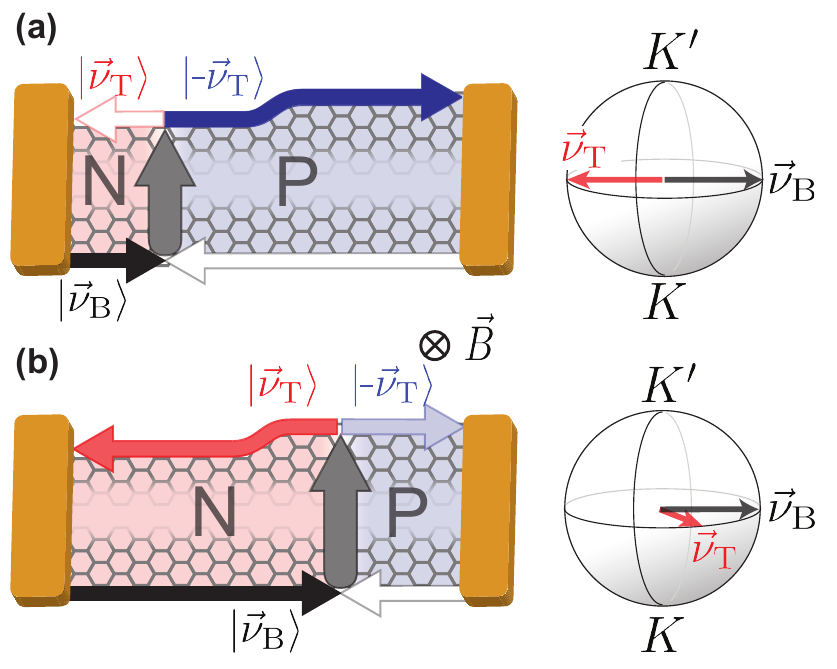}
    \caption{\textbf{Valley-isospin-dependent conductance of a (simplified) two-terminal \textit{p-n} junction at high magnetic fields. a,} Charge carriers are injected to the bottom-edge and guided along the \textit{p-n} junction to the top-edge. If at the position of the \textit{p-n} junction the relative angle between the valley-isospins at the two edges (same polarity) is equal to $\pi$, back-reflection is forbidden. The valley-isospin configuration for bottom- (black, $\vec{\nu}_{\text{B}}$) and top-edge (red, $\vec{\nu}_{\text{T}}$) is illustrated on the right-hand side. \textbf{b,} By moving the \textit{p-n} junction to a region of the flake with a different width, the relative angle between $\vec{\nu}_{\text{B}}$ and $\vec{\nu}_{\text{T}}$, which is denoted by \si{\Phi}, can change. For \si{\Phi}$ \neq \pi$ a non-zero back-reflection is allowed. In the experiment multiple steps are present.}
    \label{fig:Concept}
\end{figure}

At high magnetic fields the flow of charge carriers in graphene is restricted to quantum Hall channels moving along the edges while the bulk is insulating \cite{Zhang05,Novoselov05,Klimov15}. In the presence of a \textit{p-n} interface the charge carriers additionally flow along the \textit{p-n} junctions within the bulk \cite{Abanin07,Williams09}.
To produce a valley filter using quantum Hall states, a two-terminal graphene nanoribbon (GNR) with a \textit{p-n} junction being smooth on the atomic scale and located perpendicular to the transport direction was suggested \cite{Tworzydlo07}, as sketched in Fig.~\ref{fig:Concept}. If only the lowest Landau level (LLL) is occupied, the conductance is exclusively determined by the width and the chirality of the GNR in the vicinity of the \textit{p-n} junction \cite{Tworzydlo07}. A conductance of $G=$\SI{2}{e^2/h} corresponds to the full transmission of the spin-degenerate edge-channel of the LLL which was injected at the left contact, guided along the \textit{p-n} junction and then transmitted to the right contact, as sketched in Fig.~\ref{fig:Concept}a. On the other hand, a value smaller than \SI{2}{e^2/h} implies a finite back-reflection, as sketched in Fig.~\ref{fig:Concept}b. This set-up is analogous to a spin valve \cite{Tombros07}: an isospin valve for which the bottom and the top edges are valley polarizer and analyzer units with the \textit{p-n} interface playing the role of the channel.
The valley degree of freedom can be described with a two-component spinor
wave function with quantization axis $\vec{\nu}$, which can be represented on the Bloch sphere, where the north- and south-pole of the sphere represent the $K'$ and $K$ valley, respectively.
In the LLL the sublattice (A or B atoms) and valley degree of freedom ($K$ and $K'$) are directly coupled to each other \cite{Zheng02}. Since for a zigzag GNR the top (bottom) edge consists of A (B) atoms exclusively, the valley-isospin is pointing to the $K$ ($K'$) valley. On the other hand, both edges of an armchair GNR consist of A and B atoms each. Thus the valley-isospin lies on the equator of the Bloch-sphere, which is a coherent superposition of the two valleys \cite{Beenakker08}. In a perfect armchair GNR, it is the relative angle \si{\Phi} between the two valley-isospins at the two opposite edges  that determines the conductance through the device \cite{Tworzydlo07}:
\begin{equation}\label{eq_general}
G=\frac{e^2}{h}\left(1-\cos \si{\Phi} \right),
\end{equation}
where \si{\Phi} depends on the number of unit cells ($N$) between bottom- and top-edge of the GNR (see Supporting information). Therefore, clean samples without inter-valley scattering along the \textit{p-n} junction are essential to observe an isospin-dependent conductance. Similar width-dependent oscillations are as well expected for zigzag GNRs \cite{Tworzydlo07}.
Even though considerable effort has been invested on theoretical studies \cite{Akhmerov07,Tworzydlo07,Low09,Fraessdorf16}, an experimental proof remains missing. The latter can be attributed to diffusive transport and lack of control over the edge structure.
Here we combine a position-tunable \textit{p-n} interface with non-uniform edges in a ballistic device. The idea of our experiment is illustrated in Fig.~\ref{fig:Concept}, where an ideal armchair GNR with its width changing only at a single position is sketched.
The valley-isospin configuration of the bottom- and top-edge in the vicinity of the \textit{p-n} junction (valley-polarized along $\vec{\nu}_{\text{B}}$ and $ \vec{\nu}_{T}$ respectively, plotted at the same polarity of the junction) is shown for two different situations. Depending on the exact position of the \textit{p-n} junction, which can be shifted as a function of electrostatic gate-voltages \cite{Handschin16}, the edge polarization $\vec{\nu}_{\text{T}}$ varies (while $\vec{\nu}_{\text{B}}$ remains fixed), resulting in a different conductance as shown in Fig.~\ref{fig:Concept}a and Fig.~\ref{fig:Concept}b. By moving the \textit{p-n} junction we are able to locally probe the valley isospin configurations at different positions of the edge. We report on conductance oscillations appearing in the presence of a  \textit{p-n} junction and in the regime of the LLL.  The oscillating conductance is observed to depend on the position of the \textit{p-n} interface and agrees with quantum transport simulations. In contrast to former studies of diffusive, two-terminal \textit{p-n} junctions in graphene \cite{Williams07},
where the conductance is dominated by mode mixing \cite{Abanin07}, we enter a novel regime where the conductance is dominated by valley-isospin physics. Surprisingly, even though our samples have rough edges the conductance oscillations are still large in the order of \si{e^2/h}.\\


\begin{figure}[tb]
    \centering
      \includegraphics[width=1\columnwidth]{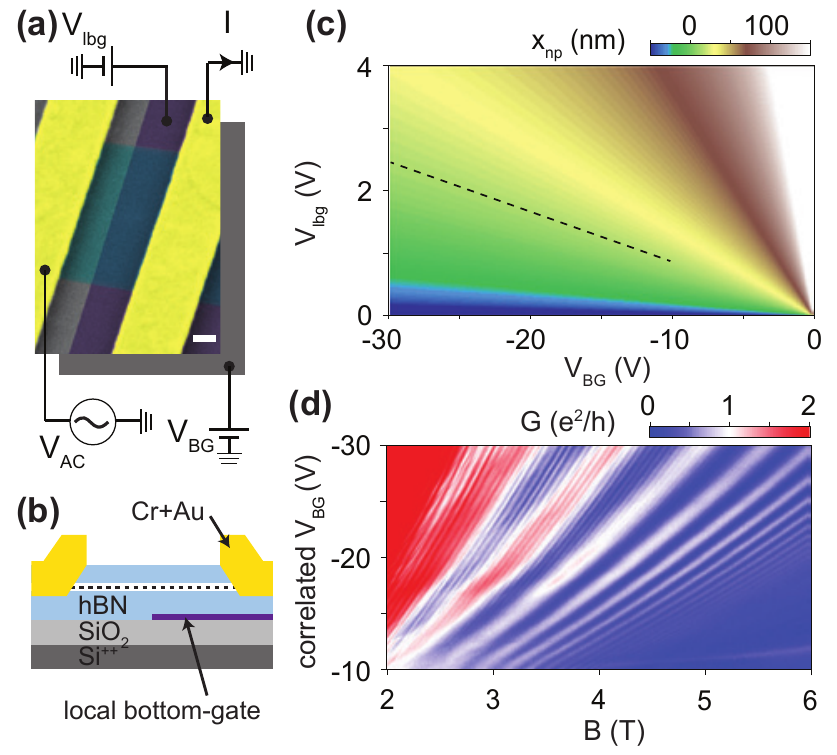}
    \caption{\textbf{Experimental set-up and basic characterization. a,} False-color SEM image where the leads are colored yellow, the graphene encapsulated in hBN is colored cyan and the local bottom-gate is colored purple. Scale-bar equals $200\,$nm. \textbf{b,} Schematic side-view of a two-terminal \textit{p-n} junction as shown in (a). \textbf{c,} Simulated \textit{p-n} interface position $x_\text{pn}$ as a function of $V_{\text{BG}}$ and $V_{\text{lbg}}$ for the device geometry. \textbf{d,} Magnetoconductance oscillations in the bipolar regime (with a $pn$-junction present) is a characteristic signature of high graphene quality. The conductance $G$ was measured along the linecut indicated in (c) with a black, dashed line.}
    \label{fig:Exp_Setup}
\end{figure}

A false-color SEM image of an encapsulated graphene \textit{p-n} junction device is shown in Fig.~\ref{fig:Exp_Setup}a.
The hBN/graphene/hBN heterostructure were assembled following the dry pick-up technique described in Ref.~\cite{Wang13}. The full heterostructure was transferred onto a pre-patterned few-layer graphene used as local bottom-gates. Standard e-beam lithography was used to define the Cr/Au side-contacts with the bottom hBN layer (\SI{70}{nm} in thickness) not fully etched through in order to avoid that the leads are shorted to the underneath lying bottom-gates.
The graphene samples were shaped into \SI{1.5}{\micro m} wide channels using a CHF$_{3}$/O$_{2}$ plasma. For more details, see Supporting information.
The charge-carrier mobility $\mu$ was extracted from field effect measurements yielding $\mu$\SI{\sim 80000}{cm^2 V^{-1} s^{-1}}.
The \textit{p-n} junction is formed by a global back- and a local bottom-gate (shown in Fig.~\ref{fig:Exp_Setup}a,b) which allows for an independent tuning of the doping on each side of the \textit{p-n} junction. With the above given device geometry, the potential profile of the \textit{p-n} junction is well within the smooth limit with respect to the lengthscale of the lattice constant (see Supporting information for electrostatic simulation), a basic requirement to observe the valley-isospin dependent oscillations \cite{Tworzydlo07}.
\begin{figure*}[tb]
    \centering
      \includegraphics[width=2\columnwidth]{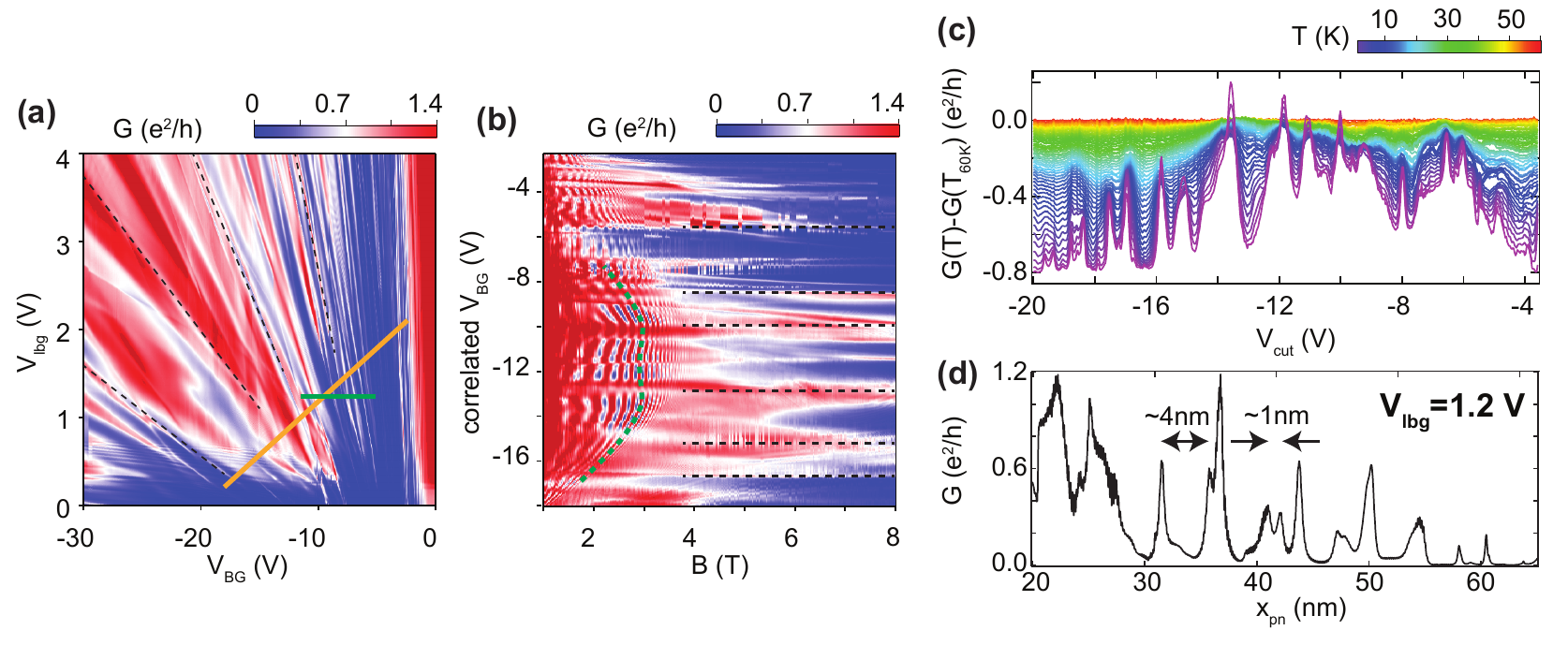}
    \caption{\textbf{Experimental results. a,} Conductance as a function of global back-gate ($V_{\text{BG}}$) and local bottom-gate ($V_{\text{lbg}}$) at $B=$\SI{8}{T}. The valley-isospin oscillations emerge as radial fringes converging to the common CNP. The black, dashed lines are a guide to the eye showing the oscillation maximum for selected valley-isospin oscillations. \textbf{b,} Linecut as indicated with the orange line in (a) as a function of magnetic field. The $x_{\text{pn}}$ dependent conductance oscillations are independent of magnetic field, and persist down to fields of roughly $B=$\SI{2}{T}. For magnetic fields below $B$\SI{\sim 3}{T} snake states appear as curved features (green, dashed line). \textbf{c,} Temperature dependence of the linecut indicated with the orange line in (a) where a background ($G(T=60\;\text{K})$) was subtracted. \textbf{d,} Linecut as indicated with the green arrow in (a) rescaled to $x_{\text{pn}}$ using the correlation between $x_{\text{pn}}$ and ($V_{\text{BG}}$,$V_{\text{lbg}}$) as given in Fig.~\ref{fig:Exp_Setup}b. For details see Supporting information.}
    \label{fig:Measurements}
\end{figure*}
The position of the \textit{p-n} interface $x_{\text{pn}}$ is adjustable due to capacitive crosstalk of the gates and is determined by the ratio of $n_{\text{lbg}}/n_{\text{BG}}$ \cite{Rickhaus15_3,Handschin16}, where $n_{\text{lbg}}$ ($n_{\text{BG}}$) is the charge carrier density tuned by the local bottom-gate (global back-gate).
An electrostatic simulation of $x_{\text{pn}}$ as a function of $V_{\text{BG}}$  and $V_{\text{lbg}}$ is shown in Fig.~\ref{fig:Exp_Setup}c, where lines of a constant $x_{\text{pn}}$ are fanning out linearly from the global charge-neutrality point (CNP). The devices show clear signatures of ballistic transport, namely Fabry-P\'{e}rot oscillations  \cite{Campos12b,Grushina13,Rickhaus13,Varlet14,Calado15,Shalom16,Handschin16}
and snake states \cite{Rickhaus15,Taychatanapat15}, which prove the absence of inter-valley scattering within the bulk of graphene. In Fig.~\ref{fig:Exp_Setup}d snake states appear as magnetoconduction oscillations as a function of correlated $V_{\text{BG}}$ (charge carrier density) and perpendicular magnetic field $B$. A detailed description of the snake states can be found in Refs.~\cite{Milovanovic14,Rickhaus15,Taychatanapat15,Kolasinski17}.\\
We now concentrate on the regime of small filling factors. A conductance map at $B=$\SI{8}{T} is shown in Fig.~\ref{fig:Measurements}a as a function of the two gate-voltages in the bipolar regime. Most prominent are the conductance oscillations on the order of \si{e^2/h}, fanning out linearly from the common CNP. The linear dependence on the two gates implies that the conductance is determined by the position of the \textit{p-n} junction (see Fig.~\ref{fig:Exp_Setup}c).
In Fig.~\ref{fig:Measurements}b a linecut (orange line in Fig.~\ref{fig:Measurements}a) is shown as a function of $B$. By tuning the gates, $x_{\text{pn}}$ can be shifted by several tens of nanometers. The conductance oscillation fringes (dashed, black lines) remain mostly unchanged (horizontal) over a large range of $B$ and thus depend only on $x_{\text{pn}}$. With decreasing magnetic field the oscillation peaks become wider and remain visible as long as edge-states are present (see Supporting information). This is in qualitative agreement with what is expected for conductance oscillations originating from valley-isospin physics since it is the valley-isospin in the range of the magnetic length $l_{\text{B}}\sim$\SI{25.6}{nm/\sqrt{B \; [T]}} which matters \cite{Tworzydlo07}.
Temperature dependence of the oscillations is shown in Fig.~\ref{fig:Measurements}c. We observe an increase of the oscillation amplitude up to the order of \si{e^2/h} while approaching the base-temperature of \SI{1.6}{K}. At high temperatures the conductance saturates close to \si{e^2/h} as expected from equilibration of the LLL's (see Supporting information).


In summary, we see conduction oscillations with an amplitude on the order of \si{e^2/h}, which are independent of doping and magnetic field to a wide extent,
but depend on $x_{\text{pn}}$. The latter suggests that the conductance is determined by the edge configuration and therefore by the local isospin configuration of the edges.
At this point we emphasize that the theory of the valley-isospin oscillation is based on the fact that only the LLL is occupied, while in part of  the measured gate-range Landau levels with $|\nu|> 2$ are populated. However, at high magnetic fields higher Landau levels remain decoupled from the LLL, and consequently do not play a role in the transport, due to the smoothness of the \textit{p-n} junction. This is in agreement with temperature-dependent measurements (see Supporting information), other transport experiments \cite{Klimov15} and theoretical studies \cite{LaGasse16}.

Using Fig.~\ref{fig:Exp_Setup}c the oscillations can be plotted as a function of $x_{\text{pn}}$ as shown in Fig.~\ref{fig:Measurements}d.
This gives us information on the edge-disorder correlation length, and  we deduce a characteristic spacing between peaks on the order of a few nanometers (with $l_{\text{B}}\sim$\SI{9}{nm}). It comes as a surprise that even for edges defined by reactive ion etching, which are expected to be rough, the conductance oscillations do not fully average out \cite{Low09}. This is supported by quantum transport simulations.\\


\begin{figure*}[tb]
    \centering
      \includegraphics[width=2\columnwidth]{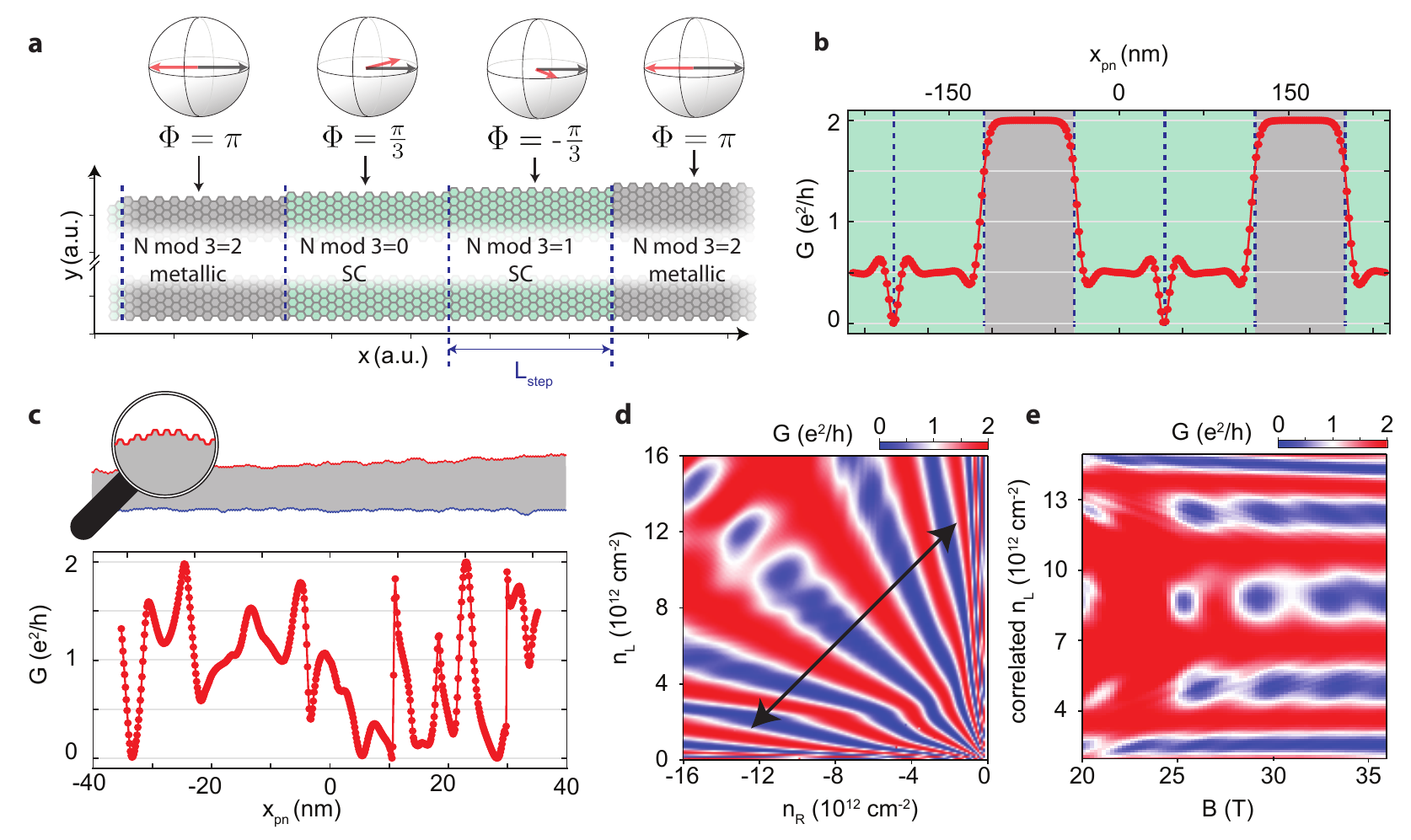}
    \caption{\textbf{Quantum transport simulations for armchair GNRs. a,} Illustration of the model used for the calculation shown in \textbf{b,} where the conductance as a function of $x_{\text{pn}}$ for a ribbon with $L_{\text{step}}\gg l_{\text{B}}$ ($L_{\text{step}}$ defined in (a)) and magnetic field $B=$\SI{16}{T} is shown. Metallic sections of the GNR are shaded in black, and semiconducting (SC) ones in green. \textbf{c,} A comparable calculation as in (b) but with disordered edges. The real-space structure of the disordered edge is shown on top of the graph (same scale in the horizontal axes). \textbf{d,} Calculated conductance of another ribbon (see text) as a function of left and right densities, $G(n_{\text{R}},n_{\text{L}})$, at $B=$\SI{36}{T}. The black arrow indicates the density sweep considered in \textbf{e,} where the conductance as a function of the correlated $n_{\text{L}}$ and $B$ is shown.}
    \label{fig:Theory}
\end{figure*}

In the following transport simulations, we focus on armchair GNRs. For zigzag GNRs and details of the calculations, see the Supporting information. In contrast to the experiment, the exact edge-profile of the GNR is known, allowing us to draw a direct relation between the edge-profile and the conductance. We consider non-parallel edges with one side flat and one side tilted (flat regions followed by a change of the ribbon width by one row of atoms, as shown in Fig.~\ref{fig:Theory}a). The resulting conductance $G$ will be solely determined by the position of the \textit{p-n} junction $x_{\text{pn}}$. Figure~\ref{fig:Theory}b shows $G(x_{\text{pn}})$ for a ribbon with width $W$\SI{\sim 40}{nm} at a constant magnetic field $B=$\SI{16}{T}. Note that the rather strong $B$ (corresponding to $l_\text{B}$\SI{\sim 4.3}{nm}) is considered here to ensure $l_\text{B} \ll W$ as in our experiment. If the \textit{p-n} junction is tuned far away from the transition regions, where $\Phi$ is $\pi$ or $\pm \pi/3$, the conductance shows plateaus with 0.5 or \SI{2}{e^2/h}. However, by approaching a transition region, the conductance $G(x_{\text{pn}})$ undergoes a smooth transition between the conductance plateaus, deviating from 0.5 or \SI{2}{e^2/h} \cite{Sekera17}. Such a regular pattern of $G(x_{\text{pn}})$, confirming the theory predicted in Ref.~\cite{Tworzydlo07}, naturally randomizes when the considered ribbon is edge-disordered. However, oscillations with high amplitudes remain, although they become irregular, as the example in Fig.~\ref{fig:Theory}c shows. The distance between neighbouring peaks is on the order of the magnetic length, similar to the experiment in Fig.~\ref{fig:Measurements}d.\\
To qualitatively reproduce the features reported in our experiment, we consider a clean and edge-disorder-free ribbon ($W$\SI{\sim 80}{nm} and $L$\SI{\sim 100}{nm}) with its charge carrier density individually tunable in the left ($n_{\text{L}}$) and right ($n_{\text{R}}$) region. The slope of the tilted edge is such that the chirality changes about every \SI{2}{nm}, and the \textit{p-n} junction shifts a few tens of \si{nm}. With a fixed magnetic field $B=$\SI{36}{T}, Fig.~\ref{fig:Theory}d shows a radial oscillation pattern fanning out from the common charge-neutrality point of $G(n_{\text{R}},n_{\text{L}})$, similar to Fig.~\ref{fig:Measurements}a. Finally, we examine the $B$ dependence of the conductance along the density sweep indicated by the black arrow in Fig.~\ref{fig:Theory}d. The horizontal fringes clearly visible in Fig.~\ref{fig:Theory}e indicate the independence of the conductance on $B$, similar to our measurement reported in Fig.~\ref{fig:Measurements}b. Note that within the density range marked in Fig.~\ref{fig:Theory}d and considered in Fig.~\ref{fig:Theory}e, the position of the \textit{p-n} junction shifts by about \SI{30}{nm}, covering about 15 steps and hence 5 periods of the alternating edge chirality, well agreeing with the number of the observed fringes shown in Figs.~\ref{fig:Theory}d and e.
The calculations on zigzag GNRs reveal comparable results as presented in Fig.~\ref{fig:Theory}b,c, which are in agreement with the results expected from ideal \cite{Tworzydlo07} and disordered GNRs \cite{Low09}.

In summary, we have shown first signatures of conductance oscillations originating from the local isospin configuration of the edges of a graphene flake. Although the edge of the flake is not controlled, the conductance is still defined by the local properties of the edges and the local width of the flake, in agreement with transport simulations. Furthermore, we can exclude that the equilibration between edge-channels at the intersection between \textit{p-n} junction and the graphene edges (so called hot-spots) is responsible for the conductance oscillations (see Supporting informations). We have observed similar oscillation in more than 15 \textit{p-n} and \textit{p-n-p} junctions,  some of them also having naturally cleaved graphene flakes \cite{Geim07} (presumably less edge-defects compared to reactive ion etching), and also on suspended \textit{p-n} junctions (see Supporting informations).
Finally, there are new techniques appearing, such as hydrogen-plasma etching \cite{Yang10,Shi11,Dobrik13} or chemical synthesis of GNRs \cite{Cai10}, allowing for a much better control over the edges. This could be used in further studies to draw a correlation between transport measurement and the edge of the measured samples (e.g. via atomic resolution imaging) underlining the isospin origin of these oscillations.



\textbf{Acknowledgments}\\
This work was funded by the Swiss National Science Foundation, the Swiss Nanoscience Institute, the Swiss NCCR QSIT, the ERC Advanced Investigator Grant QUEST, ISpinText FlagERA network and the EU flagship project graphene. M.-H.L. and K.R.  acknowledge  financial  support  by  the  Deutsche Forschungsgemeinschaft (SFB 689 and RI 681/13). Growth of hexagonal boron nitride crystals was supported by the Elemental Strategy Initiative conducted by the MEXT, Japan  and  JSPS  KAKENHI  Grant  Numbers  JP26248061, JP15K21722,  and  JP25106006. The  authors  thank Rakesh Tiwari, Anton Akhmerov, Endre T\'{o}v\'{a}ri and Mirko Rehmann for fruitful discussions.\\

\bibliographystyle{PRL}

\begin{thebibliography}{10}

\bibitem{Zhang05}
Y.~Zhang, Y.-W. Tan, H.~L. Stormer, and P.~Kim,
\newblock Nature {\bf 438}, 201 (2005).

\bibitem{Novoselov05}
K.~S. Novoselov {\em et~al.},
\newblock Nature {\bf 438}, 197 (2005).

\bibitem{Klimov15}
N.~N. Klimov {\em et~al.},
\newblock Phys. Rev. B {\bf 92}, 241301 (2015).

\bibitem{Abanin07}
D.~A. Abanin and L.~S. Levitov,
\newblock Science {\bf 317}, 641 (2007).

\bibitem{Williams09}
J.~R. Williams, D.~A. Abanin, L.~DiCarlo, L.~S. Levitov, and C.~M. Marcus,
\newblock Phys. Rev. B {\bf 80}, 045408 (2009).

\bibitem{Tworzydlo07}
J.~Tworzyd\l{}o, I.~Snyman, A.~R. Akhmerov, and C.~W.~J. Beenakker,
\newblock Phys. Rev. B {\bf 76}, 035411 (2007).

\bibitem{Tombros07}
N.~Tombros, C.~Jozsa, M.~Popinciuc, H.~T. Jonkman, and B.~J. van Wees,
\newblock Nature {\bf 448}, 571 (2007).

\bibitem{Zheng02}
Y.~Zheng and T.~Ando,
\newblock Phys. Rev. B {\bf 65}, 245420 (2002).

\bibitem{Beenakker08}
C.~W.~J. Beenakker,
\newblock Rev. Mod. Phys. {\bf 80}, 1337 (2008).

\bibitem{Akhmerov07}
A.~R. Akhmerov and C.~W.~J. Beenakker,
\newblock Phys. Rev. Lett. {\bf 98}, 157003 (2007).

\bibitem{Low09}
T.~Low,
\newblock Phys. Rev. B {\bf 80}, 205423 (2009).

\bibitem{Fraessdorf16}
C.~Fr\"{a}ssdorf, L.~Trifunovic, N.~Bogdanoff, and P.~W. Brouwer,
\newblock Phys. Rev. B {\bf 94}, 195439 (2016).

\bibitem{Handschin16}
C.~Handschin {\em et~al.},
\newblock Nano Lett. {\bf 17}, 328 (2016).

\bibitem{Williams07}
J.~R. Williams, L.~DiCarlo, and C.~M. Marcus,
\newblock Science {\bf 317}, 638 (2007).

\bibitem{Wang13}
L.~Wang {\em et~al.},
\newblock Science {\bf 342}, 614 (2013).

\bibitem{Rickhaus15_3}
P.~Rickhaus, P.~Makk, M.-H. Liu, K.~Richter, and C.~Sch\"{o}nenberger,
\newblock Applied Physics Letters {\bf 107}, 223102 (2015).

\bibitem{Campos12b}
L.~Campos {\em et~al.},
\newblock Nat Commun {\bf 3}, 1239 (2012).

\bibitem{Grushina13}
A.~L. Grushina, D.-K. Ki, and A.~F. Morpurgo,
\newblock Appl. Phys. Lett. {\bf 102}, 223102 (2013).

\bibitem{Rickhaus13}
P.~Rickhaus {\em et~al.},
\newblock Nat Commun {\bf 4}, 2342 (2013).

\bibitem{Varlet14}
A.~Varlet {\em et~al.},
\newblock Phys. Rev. Lett. {\bf 113}, 116601 (2014).

\bibitem{Calado15}
E.~V. Calado {\em et~al.},
\newblock Nat Nano {\bf 10}, 761 (2015).

\bibitem{Shalom16}
M.~Shalom {\em et~al.},
\newblock Nat Phys {\bf 12}, 318 (2016).

\bibitem{Rickhaus15}
P.~Rickhaus {\em et~al.},
\newblock Nat Commun {\bf 6}, 6470 (2015).

\bibitem{Taychatanapat15}
T.~Taychatanapat {\em et~al.},
\newblock Nat Commun {\bf 6}, 6093 (2015).

\bibitem{Milovanovic14}
Milovanovi\'{c}, M.~Ramezani~Masir, and F.~M. Peeters,
\newblock Applied Physics Letters {\bf 105}, 123507 (2014).

\bibitem{Kolasinski17}
K.~Kolasi\'{n}ski, A.~Mre\'{n}ca-Kolasi\'{n}ska, and B.~Szafran,
\newblock Phys. Rev. B {\bf 95}, 045304 (2017).

\bibitem{LaGasse16}
S.~W. LaGasse and J.~U. Lee,
\newblock Phys. Rev. B {\bf 94}, 165312 (2016).

\bibitem{Sekera17}
T.~Sekera, C.~Bruder, E.~J. Mele, and R.~P. Tiwari,
\newblock Phys. Rev. B {\bf 95}, 205431 (2017).

\bibitem{Geim07}
A.~K. Geim and K.~S. Novoselov,
\newblock Nat Mater {\bf 6}, 183 (2007).

\bibitem{Yang10}
R.~Yang {\em et~al.},
\newblock Adv. Mater. {\bf 22}, 4014 (2010).

\bibitem{Shi11}
Z.~Shi {\em et~al.},
\newblock Adv. Mater. {\bf 23}, 3061 (2011).

\bibitem{Dobrik13}
G.~Dobrik, L.~Tapaszt\'{o}, and L.~Bir\'{o},
\newblock Carbon {\bf 56}, 332 (2013).

\bibitem{Cai10}
J.~Cai {\em et~al.},
\newblock Nature {\bf 466}, 470 (2010).

\end{thebibliography}

\end{document}


\beginsupplement

\title{Supporting information for \\ Giant valley-isospin conductance oscillations in ballistic graphene}


\author{Clevin Handschin}
\thanks{These authors contributed equally}
\affiliation{Department of Physics, University of Basel, Klingelbergstrasse 82, CH-4056 Basel, Switzerland}

\author{P\'{e}ter Makk}
\thanks{These authors contributed equally}
\email{Peter.makk@unibas.ch}
\affiliation{Department of Physics, University of Basel, Klingelbergstrasse 82, CH-4056 Basel, Switzerland}

\author{Peter Rickhaus}
\thanks{These authors contributed equally}
\affiliation{Department of Physics, University of Basel, Klingelbergstrasse 82, CH-4056 Basel, Switzerland}

\author{Romain Maurand}
\affiliation{CEA, INAC-PHELIQS, University Grenoble Alpes, F-3800 Grenoble, France}

\author{Kenji Watanabe}
\affiliation{National Institute for Material Science, 1-1 Namiki, Tsukuba, 305-0044, Japan\\}

\author{Takashi Taniguchi}
\affiliation{National Institute for Material Science, 1-1 Namiki, Tsukuba, 305-0044, Japan\\}

\author{Klaus Richter}
\affiliation{Institut für Theoretische Physik, Universit\"{a}t Regensburg, D-93040 Regensburg, Germany}

\author{Ming-Hao Liu}
\affiliation{Institut für Theoretische Physik, Universit\"{a}t Regensburg, D-93040 Regensburg, Germany}
\affiliation{Department of Physics, National Cheng Kung University, Tainan 70101, Taiwan}

\author{Christian Sch\"onenberger}
\email{Christian.Schoenenberger@unibas.ch}
\affiliation{Department of Physics, University of Basel, Klingelbergstrasse 82, CH-4056 Basel, Switzerland}

\maketitle

\tableofcontents{}
\newpage

\section{Fabrication of pn-junctions with local bottom-gates}
\begin{figure}[htbp]
    \centering
      \includegraphics[width=1\textwidth]{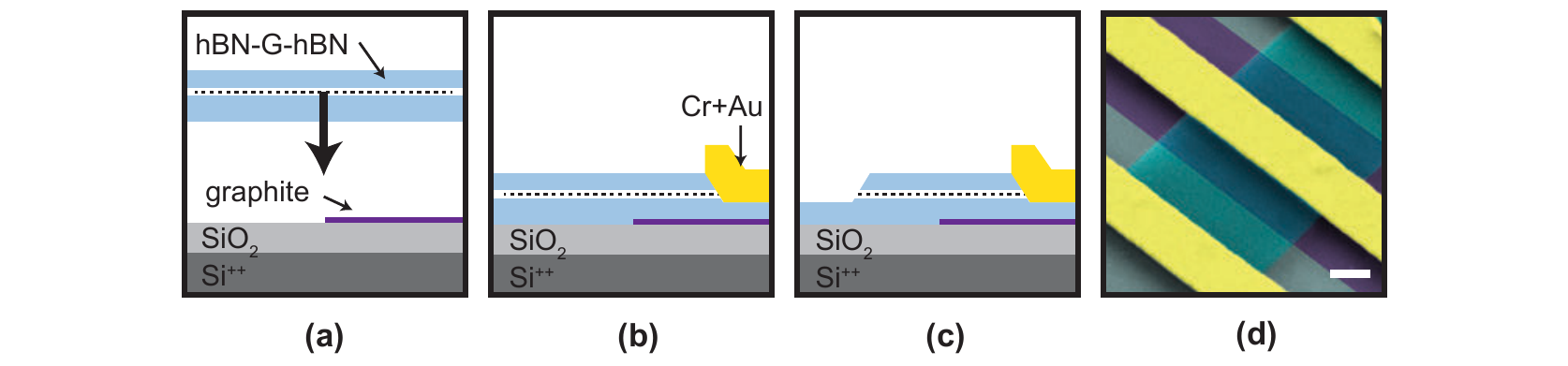}
    \caption{\textbf{Fabrication of a two-terminal \textit{p-n} junction array. a-c,} The assembly of the hBN-graphene-hBN heterostructure and establishing the side-contacts follows mostly the procedure described in \citep{Wang13}. \textbf{d,} False-color SEM image shown in (d) with bottom-gates indicated in purple. Scale-bar equals \SI{200}{nm}.}
    \label{fig:Fabricatioin}
\end{figure}

The results shown in the main-text were measured on a \textit{p-n} junction of encapsulated graphene in hexabonal boron-nitride (hBN), as illustrated in Fig.~\ref{fig:Fabricatioin}a-d. The fabrication of the hBN-graphene-hBN heterostructure follows in most steps Ref. \citep{Wang13} with some variations and extensions as explained in the following. The graphene exfoliation is done on a  Si$^{++}$/SiO$_2$ substrate (SiO$_2$ is \SI{300}{nm} thick), using the scotch-tape technique. The chips were previously cleaned using Piranha solution (98\% H$_2$SO$_4$ and 30\% H$_2$O$_2$ in a ratio of 3:1). The full heterostructure is placed on a pre-patterned few-layer graphene (using a PMMA/HSQ mask and an O$_{2}$-plasma for etching) which is later used as local bottom-gates, as shown in Fig.~\ref{fig:Fabricatioin}a. Next, self-aligned side-contacts are established to the graphene as shown in Fig.~\ref{fig:Fabricatioin}b. Self-aligned means that the same PMMA mask is used to etch down the hBN-graphene-hBN heterostructure and subsequently directly evaporate the Cr/Au (\SI{10}{nm}/\SI{50}{nm}) contacts. This leads to electrically very transparent contacts (\SI{\sim 400}{\ohm \micro m}) since the exposed graphene edge never comes into contact with any solvent or polymer. It is worth mentioning that we use cold-development ($T\sim$\SIrange{3}{5}{\degreeCelsius}) with IPA:H$_2$O (7:3) to reduce cracking of the PMMA on hBN \cite{Lee16_arxive,Rooks02}. The bottom hBN layer is not fully etched through (Fig.~\ref{fig:Fabricatioin}b), since otherwise it forms a short with the underneath lying bottom-gates. Finally, an etching step is required to define the device into \SI{1.5}{\micro m} wide channels, as illustrated in Fig.~\ref{fig:Fabricatioin}c.

\section{Ballistic transport}
\begin{figure}[htbp]
    \centering
      \includegraphics[width=1\textwidth]{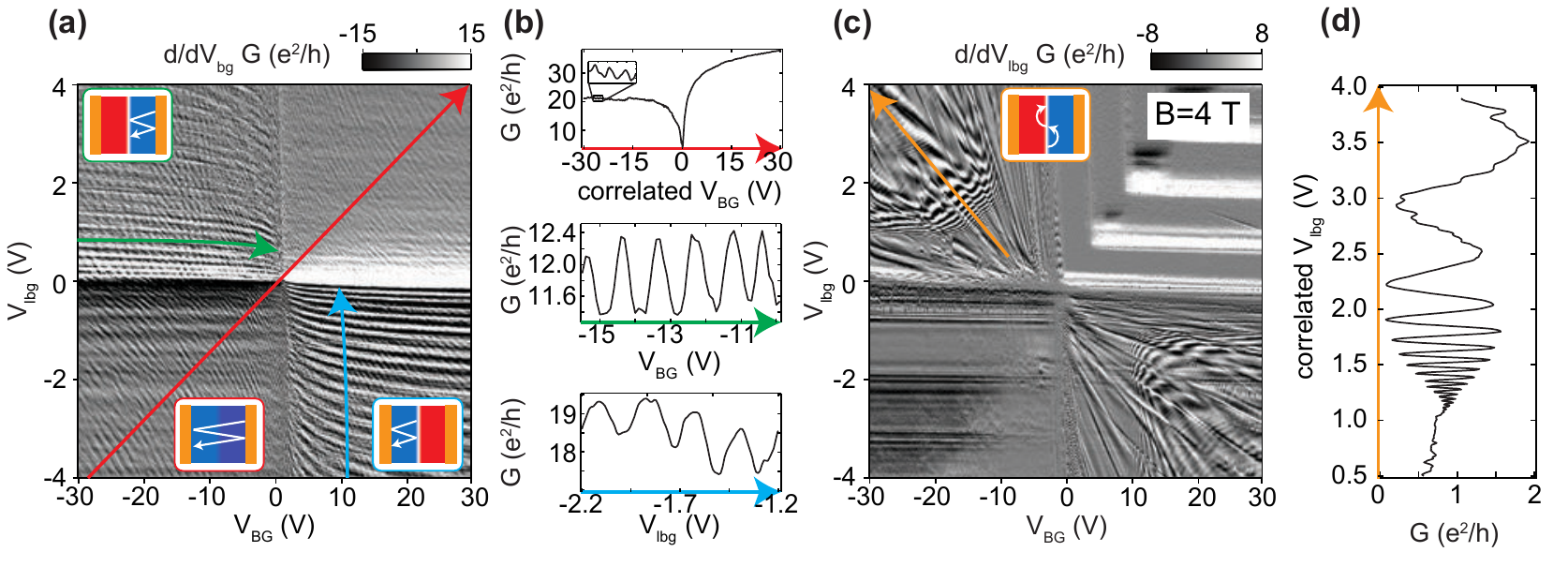}
    \caption{\textbf{Ballistic transport in a two-terminal \textit{p-n} junction. a,} Numerical derivative of the conductance as a function of global back-gate ($V_{\text{BG}}$) and local bottom-gate ($V_{\text{lbg}}$) at zero magnetic field. Fabry-P\'{e}rot oscillations in between the leads and within the right and left cavity are indicated with the red, blue and green arrows respectively. \textbf{b,} Representative linecuts within a (restricted) gate-range for of all three types of Fabry-P\'{e}rot oscillations as indicated in (a). \textbf{c,} Numerical derivative of the conductance as shown in (a), but at finite magnetic field ($B=$\SI{4}{T}). The parabolic features seen in the bipolar regimes correspond to snake states which are a clear sign of ballistic transport along the \textit{p-n} interface. \textbf{d,} Linecut in the bipolar regime as indicated in (c).}
    \label{fig:Phase_Coherence}
\end{figure}

Fabry-P\'{e}rot oscillations at zero magnetic field \cite{Campos12b,Grushina13,Rickhaus13,Varlet14,Shalom16,Calado15,Handschin16} (Fig.~\ref{fig:Phase_Coherence}a,b) or snake states at finite magnetic field \cite{Rickhaus15,Taychatanapat15}(Fig.~\ref{fig:Phase_Coherence}c,d) are signatures of ballistic transport. Ballistic transport along the \textit{p-n} junction (snake states) directly implies the absence of inter-valley scattering events between the bottom- and tog-edges, an essential ingredient to observe the valley-isospin oscillations.

\newpage

\section{Tunable position of the \textit{p-n} junction}
\begin{figure}[htbp]
    \centering
      \includegraphics[width=1\textwidth]{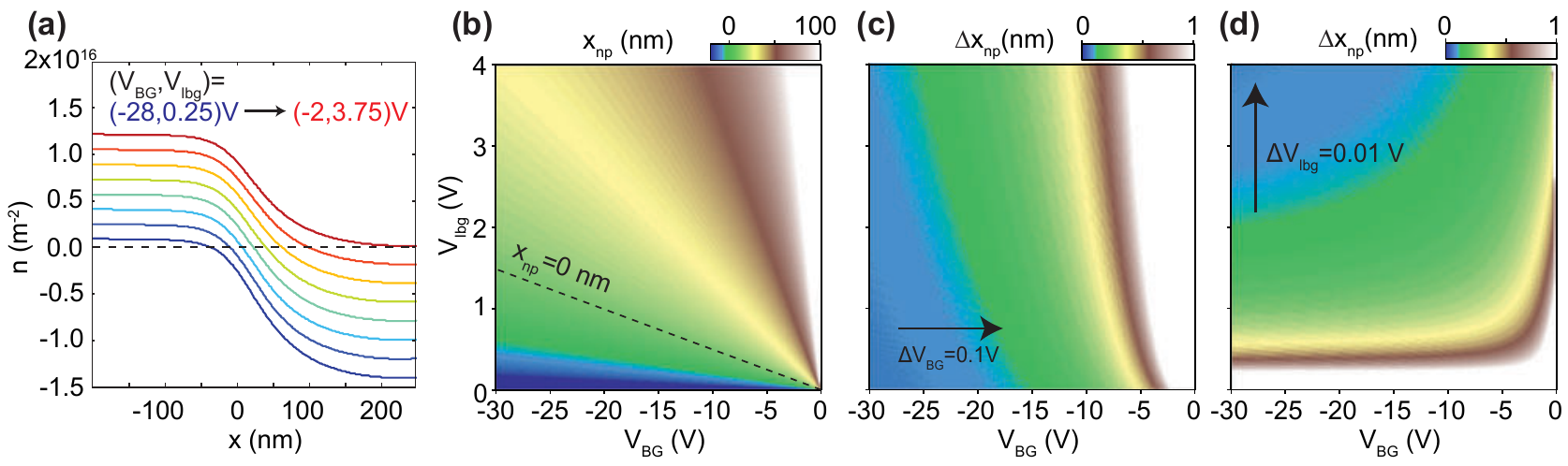}
    \caption{\textbf{Position of the \textit{p-n} junction as a function of global back-gate ($V_{\text{BG}}$) and local bottom-gate ($V_{\text{lbg}}$). a,} Exemplary density-profiles by tuning the two gates ($V_{\text{BG}},V_{\text{lbg}}$) as labelled in the header. The shift of the \textit{p-n} junction is defined as the distance between the position of the \textit{p-n} junction (zero charge carrier density, indicated with colored arrows/black, dashed lines) with respect to the edge of the local bottom-gate. \textbf{b,} The \textit{p-n} position as a function of the two gates. \textbf{c,d,} Shift of the \textit{p-n} junction for $\Delta V_{\text{BG}}=$\SI{0.1}{V} ($\Delta V_{\text{lbg}}=$\SI{0.01}{V}) while keeping $V_{\text{lbg}}$ ($V_{\text{BG}}$) fixed.}
    \label{fig:Mooving_pn}
\end{figure}
For every set of ($V_{\text{BG}}$,$V_{\text{lbg}}$) a density profile along the x-axis (defined perpendicular to the \textit{p-n} junction) can be calculated in a self-consistent way including the quantum-capacitance correction of graphene. By tuning the charge carrier densities in the two cavities of the device ($n_{\text{BG}}$, $n_{\text{lbg}}$), the position of the \textit{p-n} junction can be shifted as shown in Fig.~\ref{fig:Mooving_pn}a. In Fig.~\ref{fig:Mooving_pn}b $x_{\text{pn}}$ is plotted as a function of ($V_{\text{BG}},V_{\text{lbg}}$). How much the \textit{p-n} junction is shifted for a given stepsize ($x_{\text{pn}}/\Delta V$, where $\Delta V=$const.) is not fixed throughout the whole measurement, but depends on the doping-level in the two cavities \cite{Handschin16}, as shown in Fig.~\ref{fig:Mooving_pn}c,d. At high charge carrier doping, $x_{\text{pn}}/\Delta V$ is minimal (sub-nanometer scale for standard measurement parameters ($\Delta V_{\text{BG}}=$\SI{0.1}{V}, $\Delta V_{\text{lbg}}=$\SI{0.01}{V})
while it increases dramatically if either of the two cavities is approaching its CNP ($x_{\text{pn}}$\SI{>1}{nm} for standard measurement parameters).
\newpage

\section{Quantum transport simulations on armchair GNR}
\subsection{Conductance as a function of \textit{p-n} position $\mathbf{G(x_{\text{pn}})}$}
\begin{figure}[htbp]
    \centering
      \includegraphics[width=1\textwidth]{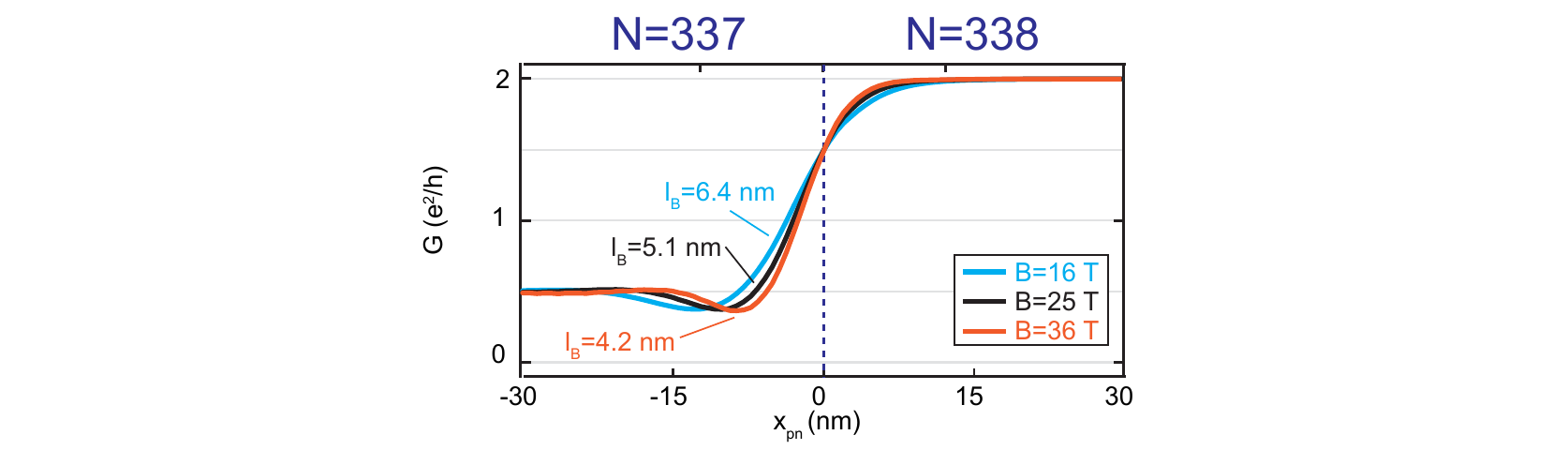}
    \caption{\textbf{Conductance as a function of $x_{\text{pn}}$ and magnetic field.} Calculations of the conductance though a GNR (with $N=337$ for $x_{\text{pn}}<0$ and $N=338$ for $x_{\text{pn}}>0$) as a function of $x_{\text{pn}}$ for different magnetic fields.}
    \label{fig:Simulation_models}
\end{figure}
In contrast to previous studies including quantum transport simulations \cite{Rickhaus15,Rickhaus15_2,Handschin16} based on scaled graphene \cite{Liu15}, we used unscaled graphene. This is crucial as we want to investigate edge-effects, which would otherwise scale correspondingly as well. In order to minimize the computational load we chose  small GNRs (\SI{\sim 40}{nm} in width and \SIrange{\sim 100}{\sim 300}{nm} in length) for the calculation presented in the following. The \textit{p-n} junction was approximated using a linear profile varying over \SI{5}{nm}. The latter revealed no qualitative change with respect to a realistic \textit{p-n} junction profile (tangens-hyperbolicus varying over \SI{\sim 15}{nm}).

The conductance though a perfect armchair GNR with a fixed number of unit cells $N$ between the two edges and a \textit{p-n} junction perpendicular to the transport direction is given by \cite{Tworzydlo07}
\begin{equation}\label{eq_general}
G=\frac{e^2}{h}\left( 1-\cos \si{\Phi} \right)
\end{equation}
where \si{\Phi} is the angle between the valley-isospin configurations of the two edges ($\vec{\nu}_{B}$ and $\vec{\nu}_{T}$). In Fig.~\ref{fig:Simulation_models} the result of a quantum-transport calculation is given for a GNR with a single-atomic step at the top-edge located at $x=0$. Far away from $x=0$ ($x_{\text{pn}}\gg l_{\text{B}}$), where $l_{\text{B}}$ is the magnetic length given by $l_{\text{B}}=25.6\;\text{nm}/\sqrt{B}$, the conductance is given by Equation~\ref{eq_general}, which corresponds to $G=$\SI{0.5}{e^2/h} or $G=$\SI{2}{e^2/h}. However, in the vicinity of $x_{\text{pn}}=0$ the conductance undergoes a smooth transition between the two adjacent conduction plateaus. An exemplary transition is shown in Fig.~\ref{fig:Simulation_models} for different magnetic fields. From the calculations shown in Fig.~\ref{fig:Simulation_models} we see that the transition length between two adjacent plateaus does depend on $l_{\text{B}}$.

\subsection{Edge disorder}
\begin{figure}[htbp]
    \centering
      \includegraphics[width=1\textwidth]{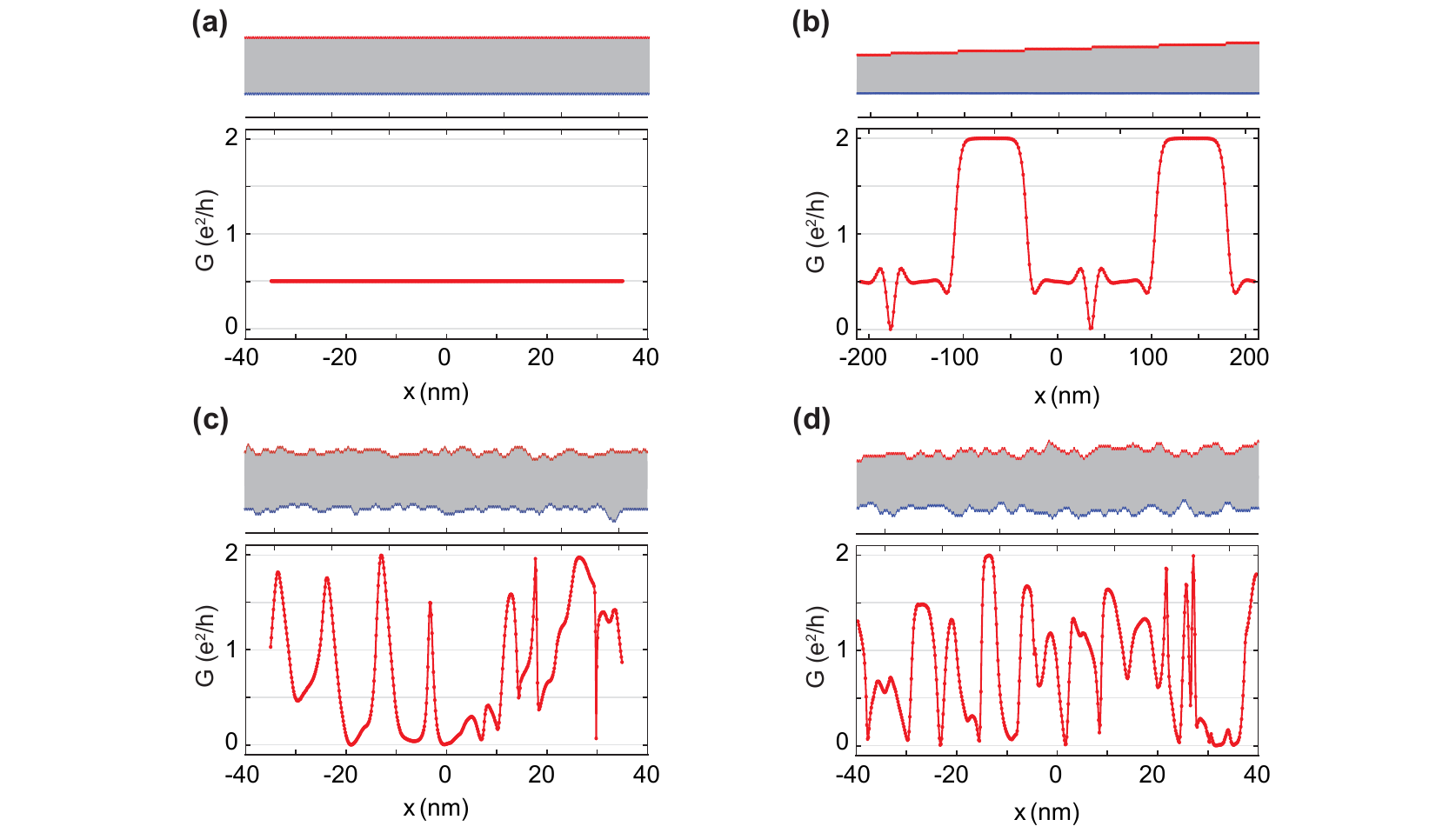}
    \caption{\textbf{Influence of random edge-roughness. a,b,} For GNRs with perfect edges and well defined steps, it is the ratio between $L_{\text{step}}$ and $l_{\text{B}}$ which defines how frequently the conductance changes. \textbf{c,d,} By introducing a random edge-roughness (a,b) become roughly equivalent since $L_{\text{step}}$ is negligible compared to the disorder length scale on which $N$ varies along the x-axis.}
    \label{fig:Simulation_edge_roughness}
\end{figure}

Having perfect armchair edges, it is $L_{\text{step}}$ which defines the distance at which the number of unit cells between bottom- and top-edge varies. Depending on the ratio of $L_{\text{step}}$ to $l_{\text{B}}$  we observe either a single plateau if there is no variation of the GNR width as shown in Fig.~\ref{fig:Simulation_edge_roughness}a, or a smooth transition between the different conduction plateaus as shown in Fig.~\ref{fig:Simulation_edge_roughness}b. In the calculations shown in Fig.~\ref{fig:Simulation_edge_roughness} the magnetic field is equal to \SI{16}{T} ($l_{\text{B}}\sim${6.5}{nm)} and $L_{\text{step}}=\infty$ ($L_{\text{step}}=$\SI{70}{nm}) respectively.
Upon introducing a random edge-roughness (details see below) , irregular conductance oscillations remain with amplitudes ranging up to \SI{2}{e^2/h}, as shown in Fig.~\ref{fig:Simulation_edge_roughness}c,d.\\


The mathematical model describing the edge-roughness includes two relevant parameters, namely the correlation length ($dL$) and the correlation amplitude ($dW$) \cite{Goodnick85,Pourfath14}. For the calculation presented in Fig.~\ref{fig:Simulation_edge_roughness} we used $dL=$\SI{1}{nm} and $dW=$\SI{4}{nm}, where the resulting edges are stabilized such that no dangling bonds remain.

\section{Quantum transport simulations on zigzag GNR}
\begin{figure}[htbp]
    \centering
      \includegraphics[width=1\columnwidth]{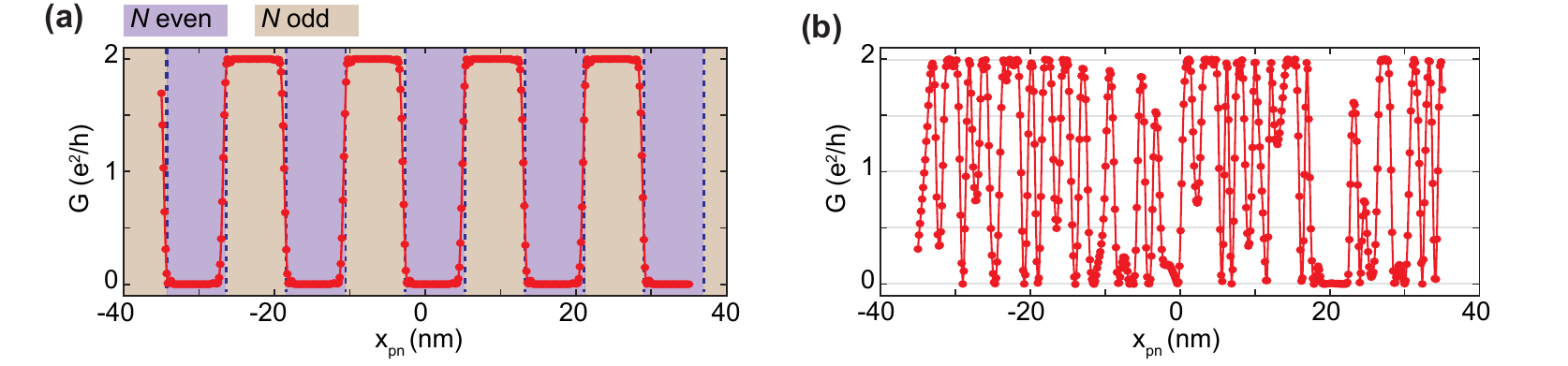}
    \caption{\textbf{Quantum transport simulations for zigzag GNRs. a,} Calculation of the conductance as a function of $x_{\text{pn}}$ similar to Fig.~4b in the main text, using the same values of $L=$\SI{100}{nm}, $W=$\SI{40}{nm}, $B=$\SI{16}{T}. \textbf{b,} Calculation as in (a) but with disordered edges, including the same correlation length ($dL=$\SI{1}{nm}) and correlation amplitude ($dW=$\SI{4}{nm}) as in Fig.~4c.}
    \label{fig:Theory_zigzag}
\end{figure}
To complement the quantum transport calculations on armchair GNRs we performed similar calculations on zigzag GNRs as shown in Fig.~\ref{fig:Theory_zigzag}. Comparable to the armchair case, the conductance takes the values of \si{0} and \SI{2}{e^2/h} far away from the transition region as shown in Fig.~\ref{fig:Theory_zigzag}a. This is in perfect agreement with the values expected from Theory \cite{Tworzydlo07}. However, in contrast to the armchair GNR the transition region in between plateaus is shorter, but remains roughly within the predicted value of $\sim l_\text{B}$.
Upon introducing edge-disorder an oscillation of the conductance results similar to the armchair case. The faster oscillations can be attributed to the shorter transition region in between plateaus as described above.

\section{Mixing of Landau levels in the bipolar regime at  $\mathbf{B=8\;}$T}
\begin{figure}[htbp]
    \centering
      \includegraphics[width=1\textwidth]{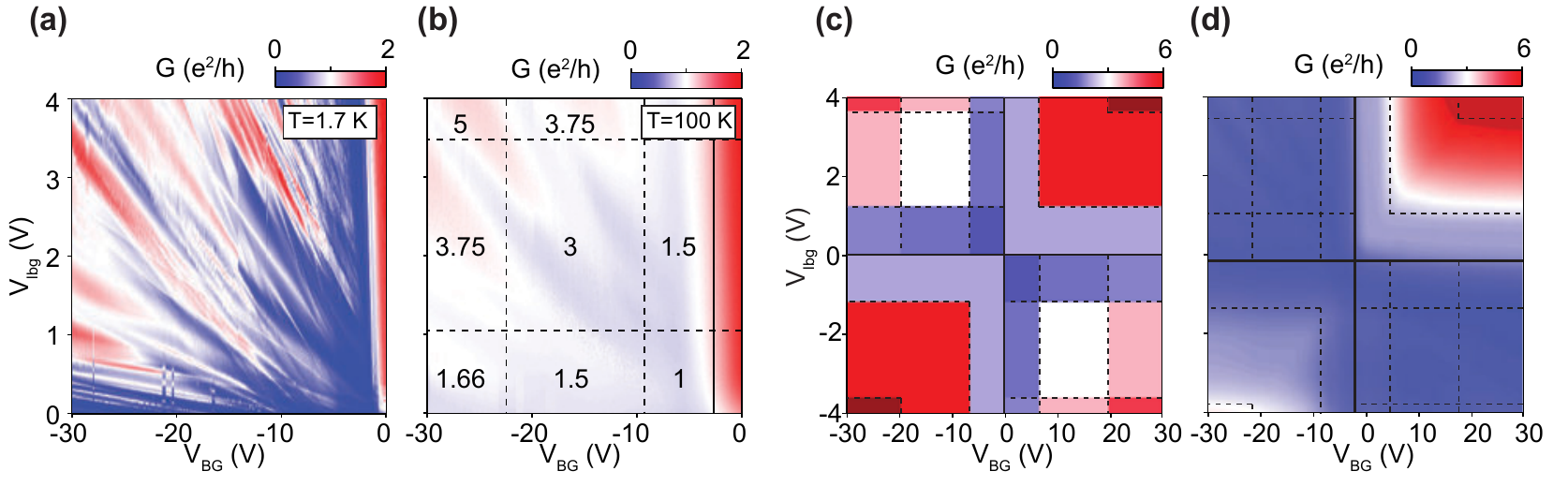}
    \caption{\textbf{Conductance in the bipolare regime at $\mathbf{B=8\;}$T for different temperatures. a,b,} With increasing temperature the valley-isospin signal becomes weaker and eventually vanishes.  At $T=$\SI{100}{\kelvin} expected conductance values, if assuming full equilibration along the \textit{p-n} interface \cite{Abanin07}, are indicated. \textbf{c,} Calculation of the conductance following Equation~\ref{eq.Mixing}. \textbf{d,} Same measurement as in (b), but for an extended gate-range. }
    \label{fig:Temp_dependence}
\end{figure}

The theory of the valley-isospin oscillation is based on the fact that p- and n-doped regions are at their lowest Landau levels ($\nu= \pm 2$, $\nu$ being the filling factor), while within the measured gate-range Landau levels with $|\nu| > 2$ are populated. However, the conduction oscillation attributed to the valley-isospin oscillations seem to persist in regions with $|\nu| > 2$ as shown in Fig.~\ref{fig:Temp_dependence}a. This suggests, that Landau levels with $|\nu|>2$ are decoupled from the LLL and do not play role in the transport. This is a result of the smoothness of the \textit{p-n} interface, which spatially separates the Landau levels, as shown in previous publications \cite{Klimov15,LaGasse16}. This claim is supported by the observation of a conductance which remains \SI{1}{e^2/h} in the bipolar regime, independent of the filling-factor, which is shown in Fig.~\ref{fig:Temp_dependence}b at elevated temperature, where the valley-isospin oscillations vanishes. This behaviour is only expected if the lowest Landau levels is decoupled from higher ones, since assuming full edge-state equilibration along the \textit{p-n} interface (within the bipolar regime) the conductance is given by \cite{Abanin07}:
\begin{equation}\label{eq.Mixing}
G=\frac{e^2}{h}\frac{|\nu_{\text{BG}} \cdot \nu_{\text{lbg}}|}{|\nu_{\text{BG}}| + |\nu_{\text{lbg}}|},
\end{equation}
where $\nu_{\text{BG}}$ and $\nu_{\text{lbg}}$ are the filling factors within the cavities tuned by either global back-gate or the local bottom-gate. The filling factors were calculated according to $\nu=(n \cdot h)/(e \cdot B)$ where the charge carrier density ($n$) is calculated assuming a plate-capacitor model using a \SI{300}{nm} thick SiO$_{2}$ and \SI{70}{nm} thick bottom-hBN. In Fig.~\ref{fig:Temp_dependence}c the expected conduction-values following Equation~\ref{eq.Mixing} are shown, with the corresponding measurement in Fig.~\ref{fig:Temp_dependence}d. The values in the hole-doped, unipolar regime deviate slightly from the calculation, since in that case the contact-doping (n-type) creates a \textit{p-n} junction in its proximity. We can therefore see that the equilibration picture is not suited to explain the measured values. The smoothness of the junction lets higher Landau levels be too far away to interact, therefore the conductance stays well below \SI{2}{e^2/h}.


\section{Temperature dependence of valley-isospin oscillations}
\begin{figure}[htbp]
    \centering
      \includegraphics[width=1\textwidth]{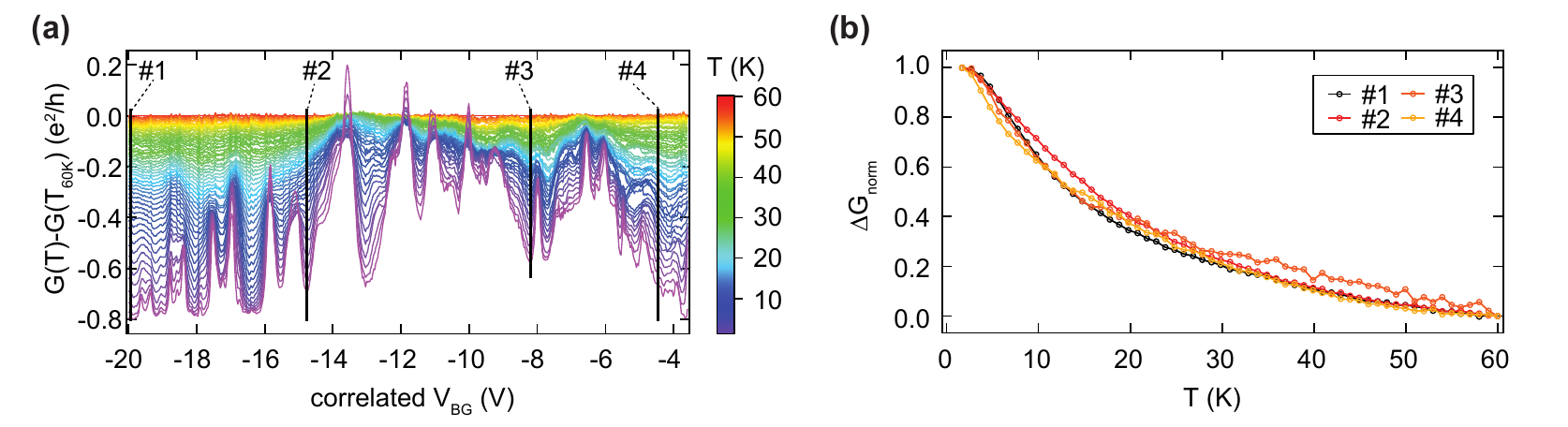}
    \caption{\textbf{Temperature dependence of valley-isospin oscillations. a,} Valley-isospin oscillations, where $\Delta G(T)$ is given by $G(T)-G(T=60\;K)$ with ($i=1,2,3,4$). \textbf{b,} Normalized amplitude as a function of temperature.}
    \label{fig:Temp_dependence_linecuts}
\end{figure}

The temperature dependence of the valley-isospin oscillations is shown in Fig.~\ref{fig:Temp_dependence_linecuts}a. In Fig.~\ref{fig:Temp_dependence_linecuts}b the normalized amplitude at various positions as indicated in (a) is shown. It can be seen that the valley-isospin oscillations persist up to temperatures of \SI{\sim 60}{K}.
\newpage

\section{Limiting factors to probe the valley-isospin oscillations}
The resolution with which it is possible to resolve edge-properties depends on several factors. Here we discuss two effects, namely
\begin{itemize}
\item the stepsize $\Delta x_{\text{pn}}$, with which we shift the position of the \textit{p-n} junction between two measurement-points.
\item the magnetic-length.  
\end{itemize}

\subsection{Stepsize $\mathbf{\Delta x_{\text{pn}}}$}
\begin{figure}[htbp!]
    \centering
      \includegraphics[width=1\columnwidth]{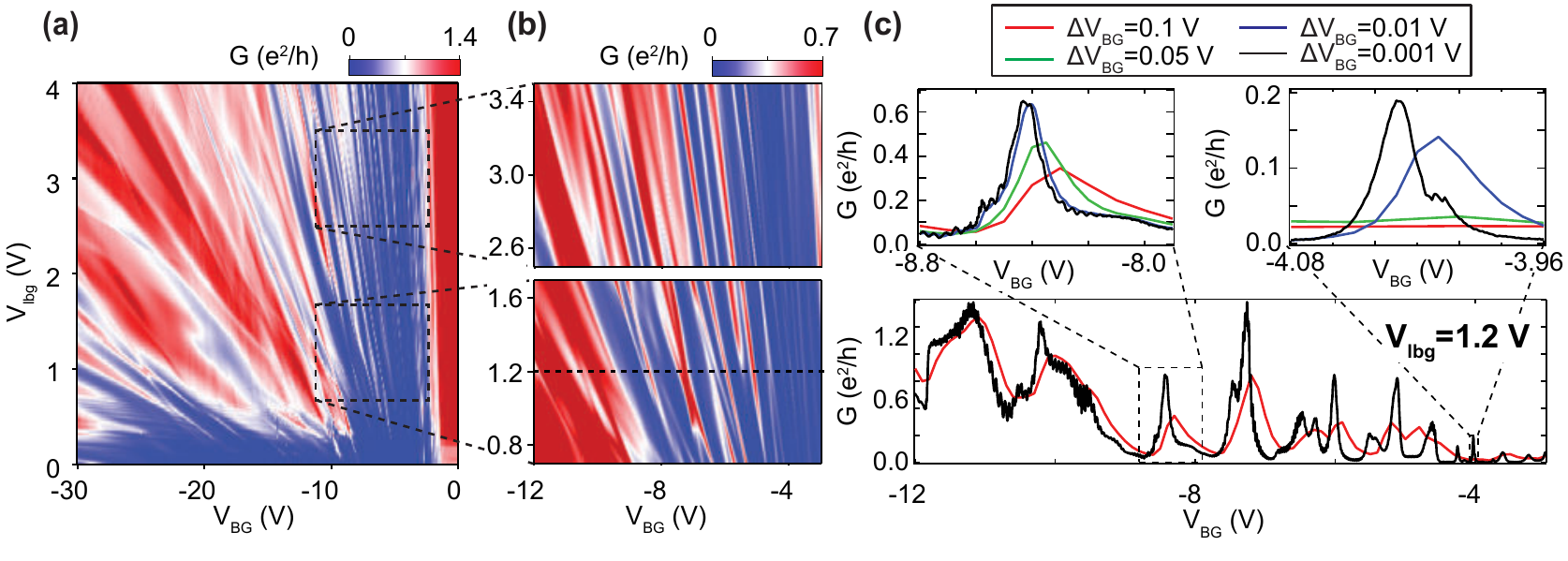}
    \caption{\textbf{Resolution limit due to the finite stepsize of $\mathbf{\Delta x_{\text{pn}}}$. a,} Conductance as a function of $V_{\text{BG}}$ and $V_{\text{lbg}}$ at $B=$\SI{8}{T}, measured with a stepsize of $\Delta V_{\text{BG}}=$\SI{0.1}{V} and $\Delta V_{\text{lbg}}=$\SI{0.01}{V}. \textbf{b,} High resolution maps ($\Delta V_{\text{BG}}=$\SI{0.02}{V} and $\Delta V_{\text{lbg}}=$\SI{0.002}{V}) for the areas as indicated in (a). \textbf{c,} Linecuts as indicated in (b) at $V_{\text{lbg}}=$\SI{1.2}{V} for different resolutions of $\Delta V_{\text{BG}}$.}
    \label{fig:SI_resolution_limit}
\end{figure}

Since features smaller than the stepsize $\Delta x_{\text{pn}}$ are washed out, it is important to tune $\Delta x_{\text{pn}}$ with sub-atomic resolution if we want to be able to resolve the edge-properties of the GNR. However, the stepsize $\Delta x_{\text{pn}}$ varies as a function of $V_{\text{BG}}$ and $V_{\text{lbg}}$ as shown in Fig.~\ref{fig:Mooving_pn}c. For a given set of ($\Delta V_{\text{BG}}$,$\Delta V_{\text{lbg}}$), the resolution is highest at high charge carrier doping while it increases dramatically as soon as either of the cavities approaches its CNP, which can be seen in Fig.~\ref{fig:SI_resolution_limit}a. In Fig.~\ref{fig:SI_resolution_limit}b the stepsizes of $\Delta  V_{\text{BG}}$ and $\Delta  V_{\text{lbg}}$ were decreased by a factor of 5, leading to an increased resolution in the areas closer to the CNP. The linecut shown in Fig.~\ref{fig:SI_resolution_limit}c was measured with different resolutions of $\Delta V_{\text{BG}}$. Far away from the Dirac peak ($V_{\text{BG}}$\SI{<-10}{V}) a resolution of $\Delta V_{\text{BG}}=$\SI{0.1}{V} seems sufficient to resolve all the peaks which are seen as well with the highest resolution  ($\Delta V_{\text{BG}}=$\SI{0.001}{V}). With $V_{\text{BG}}$ approaching the CNP only the traces with smaller stepsize of $\Delta V_{\text{BG}}$ seem to capture all the relevant features. In the very vicinity of the CNP, even $\Delta V_{\text{BG}}=$\SI{0.001}{V} seems insufficient to resolve all features.\\
At very low doping the graphene additionally enters an insulating state (plateau at $\nu=0$) which might explain as well the decrease (or even the absence) of the oscillation amplitude upon approaching a CNP. This is in contrast to the simulation where no decrease of the oscillation amplitude is observed. This is because electron-electron interaction is not taken into account, a ingredient essential for the formation of an insulating state ($\nu=0$ plateau).

\subsection{Magnetic length $\mathbf{l_{\text{B}}}$}
\begin{figure}[htbp!]
    \centering
      \includegraphics[width=1\columnwidth]{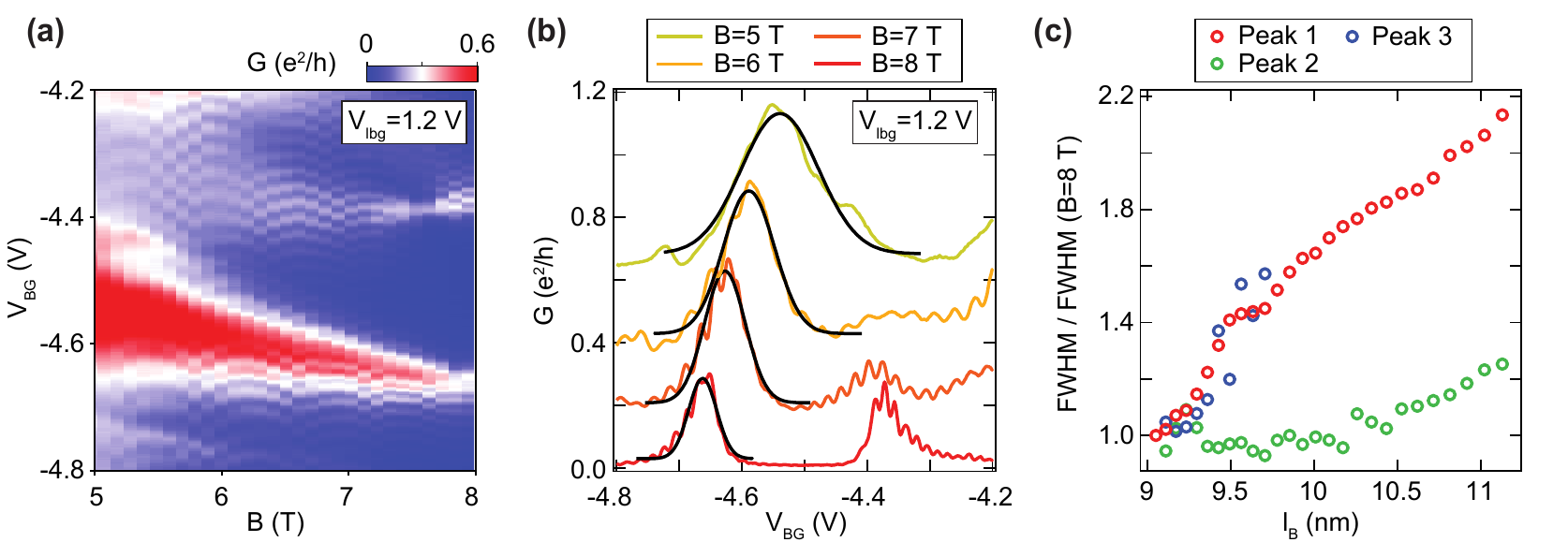}
    \caption{\textbf{Resolution limit due to the magnetic length $\mathbf{l_{\text{B}}}$. a,} Zoom at a valley-isospin oscillation as a function of magnetic field. \textbf{b,} Representative slices from (a) for different magnetic fields (slices are off-set by \SI{0.15}{e^2/h} for clarity). The width of the peaks were fitted with a Gaussian as shown in black. \textbf{c,} The evolution of the full width half maximum (FWHM), normalized by the FWHM at $B=$\SI{8}{T}, is plotted as a function of magnetic length $l_{\text{B}}$ for three individual peaks. The data from (a) and (b) corresponds to Peak 1.}
    \label{fig:SI_Peak_width_evolution}
\end{figure}

In Fig.~\ref{fig:SI_Peak_width_evolution}a the evolution of a valley-isospin oscillation is plotted as a function of magnetic field while in Fig.~\ref{fig:SI_Peak_width_evolution}b it is illustrated how a Gaussian is fitted to the peak in order to extract the full width at half maximum (FWHM). At $B=$\SI{8}{T} two individual peaks are visible and the FWHM of the peak(s) is (are) smallest. However, with increasing $l_{\text{B}}$ (decreasing magnetic field) the peaks broaden and eventually merge into one single peak, as seen for the linecut at $B=$\SI{5}{T} in Fig.~\ref{fig:SI_Peak_width_evolution}b. The normalized FWHM was plotted as a function of the magnetic length for three individual peaks, as shown in Fig.~\ref{fig:SI_Peak_width_evolution}c. Depending on the initial width of the peak (which depends on the edge-configuration), the relative change with increasing $l_{\text{B}}$ can vary significantly. Nevertheless, the FWHM increases for all of them with increasing magnetic length.\\

\section{Valley-isospin oscillations in suspended \textit{p-n} junctions}
\begin{figure}[htbp]
    \centering
      \includegraphics[width=1\textwidth]{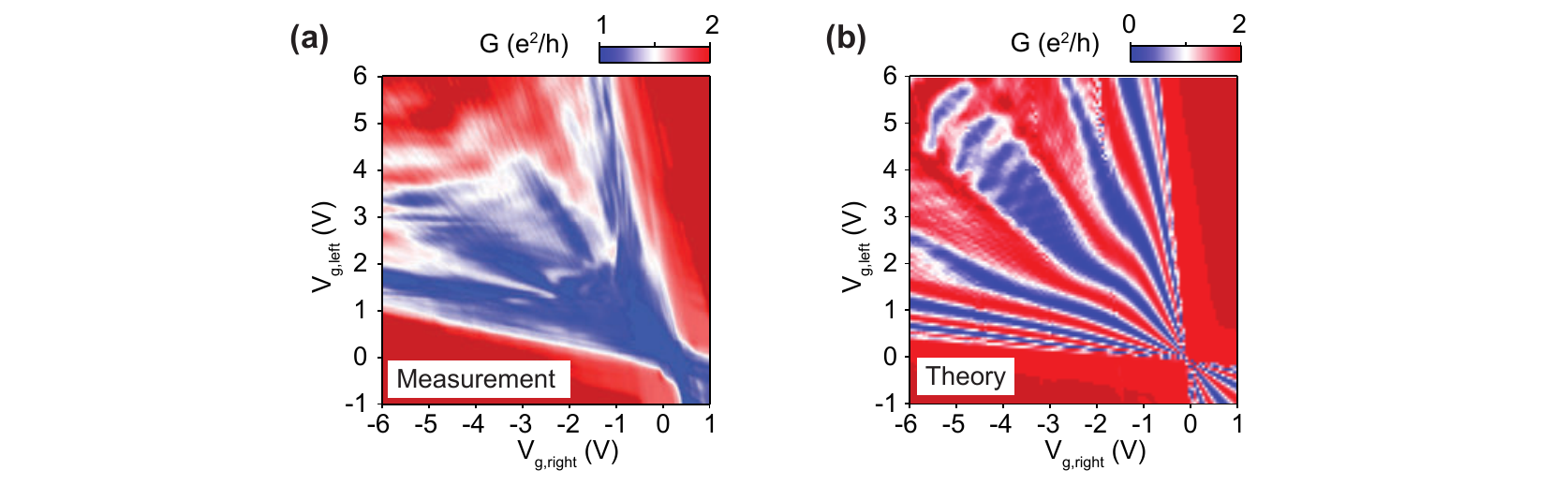}
    \caption{\textbf{Valley isospin oscillations in suspended devices at $\mathbf{B=120\;}$mT. a,} Conductance of a suspended device in the bipolar regime as a function of two local bottom-gates, which are located \SI{\sim 400}{nm} below the graphene. \textbf{b,} Simulation of a trapezoidal graphene flake (armchair edges) with scaled graphene.}
    \label{fig:SI_suspended_with_theory}  
\end{figure}

The valley-isospin oscillations were not only observed in encapsulated \textit{p-n} devices, but as well in suspended \textit{p-n} devices as shown in Fig.~\ref{fig:SI_suspended_with_theory}a. For fabrication-details of the suspended device see Ref.~\cite{Tombros11,Rickhaus13}. Quantum transport calculations on scaled graphene \cite{Liu15} in the absence (Fig.~\ref{fig:SI_suspended_with_theory}b) of bulk-disorder was performed.
\newpage

\section{Hot-spot equilibration}
Another effect which might give rise to a \textit{p-n} junction position dependent oscillation could be the probing of equilibration hot-spots along the sample edge. These hot spots could originate from increased local disorder or chemical doping.  At high magnetic field the LLL splits due to electronic interactions \cite{Zhang06,Young12,Amet14}, and instead of a single channel localized at zero doping two (or four) channels are formed on opposite side of the zero density region. These channels can equilibrate at the edges of the sample, where their mixing is possible, and by moving the \textit{p-n} junction the equilibration rates can change leading to conductance oscillation. However, since these channels are not positioned at zero charge carrier density any more, their position relative to the center of the \textit{p-n} junction is also tuned by the magnetic field. This contradicts our finding, because the position of these oscillations remains magnetic field independent. Also in the gate-gate map such oscillations might lead to more complex, hyperbolic lines \cite{Wei17_arXive}, opposite to our findings, where the lines are fanning out linearly from the common CNP. Finally, our simulations reproduce the experimental findings even in the absence of electron-electron interaction and splitting of the lowest Landau level.

\newpage

\section{Nomenclauture of zigzag and armchair GNR}
\begin{figure}[htbp]
    \centering
      \includegraphics[width=1\textwidth]{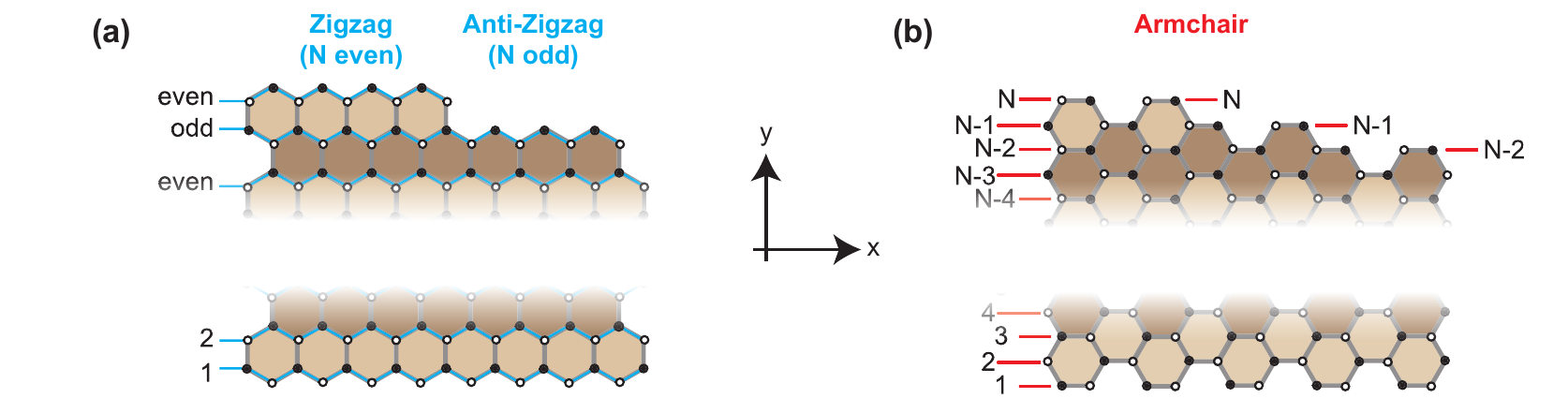}
    \caption{\textbf{Counting nomenclauture of a zigzag and armchair GNR. a,} Sketch how the number of unit cells between the two edges in y-direction are counted for zigzag (anti-zigzag) and \textbf{b,} armchair GNRs.}
    \label{fig:nomenclauture}
\end{figure}
The nomenclauture on how the number of unit cells are counted is shown in Fig.~\ref{fig:nomenclauture} for both, zigzag and armchair GNR's.\\

\textbf{Conduction plateaus for zigzag GNRs:}\\
\begin{equation}\label{eq_zigzag}
  G=\begin{cases}
    2\;e^2/h \quad \text{N odd} \quad (\si{\Phi}=0, \: \text{anti-zigzag})\\
    0\;e^2/h \quad \text{N even} \quad (\si{\Phi}=\pi, \: \text{zigzag})
  \end{cases}
\end{equation}

The two edges are coupled via the edge-channel running along the \textit{p-n} interface, which is valley degenerate \cite{Lukose07,Pereira07} because a smooth electrostatic potential (with respect to the lattice constant $a$) does not couple the valleys in the Dirac equation. Therefore in the case of $N=$odd ($G=0$) inter-valley scattering is intrinsic for zigzag edges \cite{Akhmerov08,Low09}.

\textbf{Conduction plateaus for armchair GNRs:}\\
\begin{equation}\label{eq_armchair}
  G=\begin{cases}
    2\;e^2/h \quad N\bmod 3=2 \quad (\si{\Phi}=\pi)\\
    0.5\;e^2/h \quad \text{N otherwise} \quad (\si{\Phi}=\pm \pi /3)
  \end{cases}
\end{equation}
The criterion of $N \bmod 3=2$ is equivalent for obtaining a metallic GNR. In contrast to the zigzag edge, the conductance alternates with $3N$ and the valley-isospin on both edges is a coherent superposition of the K and K' valley \cite{Beenakker08}.

\bibliographystyle{PRL}